\documentclass[prb,aps,twocolumn,amsdtx,showpacs]{revtex4-1}
\usepackage{graphicx}
\usepackage{amsmath}
\usepackage{amssymb}
\usepackage{color}

\newcommand{\bleq}{\ifpreprintsty
                   \else
                   \end{multicols}\vspace*{-3.5ex}{\tiny
                  \noindent\begin{tabular}[t]{c|}
                  \parbox{0.493\hsize}{~} \\ \hline \end{tabular}}
                   \fi}
\newcommand{\eleq}{\ifpreprintsty
                 \else
                   {\tiny\hspace*{\fill}\begin{tabular}[t]{|c}\hline
                    \parbox{0.49\hsize}{~} \\
                    \end{tabular}}\vspace*{-2.5ex}\begin{multicols}{2}
                    \fi}
\newcommand{\bcols}{\ifpreprintsty\else\begin{multicols}{2}\fi}
\newcommand{\ecols}{\ifpreprintsty\else\end{multicols}\fi}

\newcommand \beq  {\begin{equation}}
\newcommand \eeq  {\end{equation}}
\newcommand \bea {\begin{eqnarray} }
\newcommand \eea {\end{eqnarray}}

\newcommand \rarrow{\rightarrow}

\begin{document}

\title{Spin ice on the trillium lattice studied by Monte Carlo calculations}

\author{Travis E. Redpath}
\affiliation{Brandon University, Brandon, Manitoba, Canada, R7A 6A9}
\author{John M. Hopkinson}
\affiliation{Brandon University, Brandon, Manitoba, Canada, R7A 6A9}
\date{\today}

\begin{abstract}

 We study a local ferromagnetic Ising model for classical spins on the trillium lattice. The ground state of this model features two spins out(/in) and one spin in(/out) on each triangle, and leads to a macroscopic ground state degeneracy. Our Monte Carlo simulations find a ground state entropy intermediate to that of spin ice on the kagome and pyrochlore lattices, suggesting that trillium spin ice is highly frustrated. To motivate the search for trillium spin ice, we calculate the magnetic susceptibility and structure factor. We note the qualitative resemblance of the susceptibility to previously published work on EuPtSi, which features local moments on the trillium lattice.
\end{abstract} 

\maketitle

\section{Introduction}

 Recently there has been great interest in the possibility that spin ice materials may provide a route to the realization of deconfined magnetic monopoles in a fully three dimensional (3D) correlated spin system. While such proposals{\cite{cast,Jambert}} and their experimental support{\cite{Bram,Fennell}} have focused on materials based on the pyrochlore structure, a growing stable of lattice structures, and accompanying spin ice materials, is developing which may provide complementary routes to the same fundamental physics. In particular the existence of (two-dimensional) kagome spin ice was predicted in 2002 by Wills {\it{et al.}}{\cite{wills}}, extended to dipolar spin ice by Chern {\it{et al.}}{\cite{Chern}}, and recently realized with dipolar mesoscopic interactions between lithographically etched ferromagnetic islands{\cite{Yi}}. The existence of (3D) hyperkagome dipolar spin ice has recently been proposed{\cite{Carter}} as a pure limit of ``diluted'' spin ice{\cite{XK}}.

 Spin ice on the pyrochlore lattice is one of the most well understood examples of a geometrically frustrated (GF) system both theoretically{\cite{Isakov,cast,Den,Melko,siddharthan}} and experimentally{\cite{Harris,morris,ramirez,bramwell,kanada}}. GF materials have particularly interesting properties which arise because the symmetries of the triangle- or tetrahedral-based lattice structures lead to interactions between charges and/or spins which cannot simultaneously be uniquely minimized. Generically one finds many equal energy ground state configurations among which selection of a long range ordered ground state manifold proceeds via higher order processes such as order by disorder produced by finite temperature or quantum fluctuations. One of the earliest recognized frustrated materials was water, which was found to retain a finite entropy to very low temperatures by Giauque and Stout{\cite{giauque}} and explained in terms of proton disorder by Pauling{\cite{Pauling}}. A short 14 years later it was recognized that macroscopic ground state entropies could arise in models of magnetism on GF lattices when Wannier{\cite{wannier}} showed that antiferromagnetic (AFM) Ising spins on the triangular lattice had a ground state entropy per spin of 0.338314. By the mid 1980s it was recognized that a number of Ising AFMs exhibit finite ground state entropies{\cite{liebmann}} and the list of GF lattices exhibiting finite ground state entropies has continued to grow{\cite{diep}}, now including some quantum spin models in addition to the spin ice materials motivating this work.

 The term ``spin-ice'' was coined by Harris {\it{et al.}}{\cite{Harris}} to describe the physics resulting from a ferromagnetic (FM) coupling between nearest neighboring spins on the pyrochlore lattice when these local moments experienced an Ising anisotropy directed along the local $<\!111\!>$ directions (which point toward the centre of each tetrahedron). Following earlier work by Anderson{\cite{And}}, Harris {\it{et al.}} recognized that the ground state spin directions of such a local FM Ising model had a one-to-one correspondence with the displacements of hydrogen atoms from a corner-shared tetrahedral lattice in cubic water ice{\cite{note7}}. This correspondence enabled Harris {\it{et al.}} to find a simple approximation to the residual entropy of Ho$_2$Ti$_2$O$_7$ based on Pauling's estimate{\cite{Pauling}} for the residual entropy of water ice made almost 60 years earlier.

 On the pyrochlore lattice the Pauling estimate for the ground state entropy per spin of spin ice is $\frac{S}{N} = \frac{1}{2}\ln(\frac{3}{2})$. Pauling's estimate results from considering the probabilities of different ground state occurrences to be uncorrelated. To obtain this result, one notes that of the $2^N$ spin possibilities for $N$ Ising spins, only six of the sixteen possible configurations on each tetrahedron belong to the ground state with two spins in and two spins out. This reduces the degeneracy of the ground state by a factor of $(\frac{6}{16})^{N_\triangle}$ where $N_\triangle = \frac{N}{2}$ is the number of tetrahedra composing the lattice, such that for uncorrelated spins one expects a ground state degeneracy of $2^N(\frac{3}{8})^{\frac{N}{2}}$. Since Harris {\it{et al.}}'s early work it has been realized that dipolar interactions also play a strong role in the experimental realization of spin-ice materials on the pyrochlore lattice{\cite{Den,Isakov}}, with the possibility of creating an effectively FM coupling between nearest neighbor spins even when the Ising coupling has an antiferromagnetic sign.

 Wills {\it{et al.}}{\cite{wills}} extended the spin ice label to include materials based on corner-shared equilateral triangle magnetic lattices, in showing that a local FM Ising model on the kagome lattice would have an even larger residual Pauling entropy: $\frac{S}{N} = \frac{1}{3}\ln(\frac{9}{2})$. Ke {\it{et al.}}{\cite{XK}} experimentally demonstrated that the replacement of some magnetic atoms from a pyrochlore lattice with nonmagnetic atoms indeed does produce a nonmonotonic variation of the residual entropy. One of us {\cite{Carter}} has recently shown that in the disorder-free limit of such a replacement, hyperkagome spin ice, one would expect the same enhanced Pauling entropy, $\frac{S}{N} = \frac{1}{3}\ln(\frac{9}{2})$.

 In this paper, we report the first investigation{\cite{note1}} of spin ice on the trillium lattice. We show that a local FM Ising model with spins directed along the local $<\!111\!>$ axis leads to a macroscopic ground state with a residual entropy [$\frac{S}{N}\approx\ln(\frac{3}{2})$] intermediate to those of kagome and pyrochlore spin ices, indicating that this is the newest member of the spin ice class. To motivate the experimental search for trillium spin ice we calculate the magnetic susceptibility and neutron scattering structure factor for this model. We note the curious resemblance of the former quantity to susceptibility measurements of EuPtSi{\cite{adroja}}, which remains disordered to very low temperatures and whose magnetic Eu$^{2+}$ ions form a trillium lattice. We note that the neutron scattering structure factor is quite different from that of the AFM Heisenberg model{\cite{Kee}} on the trillium lattice even in the cooperative paramagnetic phase and explain why this is expected to be a generic attribute of spin ice on corner-shared equilateral triangle lattices.

\begin{figure}
\includegraphics[scale=0.34]{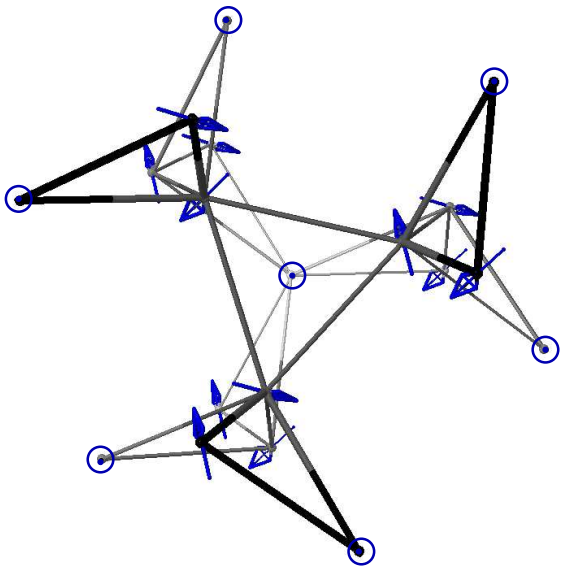}
\caption{(Color online) The Trillium lattice. The direction of each spin is shown by an arrow (the direction of the spin at one of the four sites in the unit cell is pointing out of the page). Beyond the triangle a minimum of five bonds are required for the lattice to form a closed loop.\label{figure1}}
\end{figure}

\section{Model}

 We consider a local FM Ising model on the trillium lattice with Hamiltonian,
\begin{equation}
H = J\sum_{\langle ij \rangle}\!\vec{s}_{i\text{,local}}\cdot\vec{s}_{j\text{,local}}\text{,}\label{equation8}
\end{equation}
where $\vec{s}_{i\text{,local}}=\sigma_{i\alpha}\hat{e}_\alpha$ (no sum over $\alpha$), $\sigma_{i\alpha}=\pm1$, $\langle ij \rangle$ sums over nearest neighbors, $J < 0$ is a FM coupling constant, and $\hat{e}_\alpha$ is a unit vector denoting the local easy axis of the trillium lattice as presented in Table \ref{table1}. As the dot product of any two of the nearest neighbor unit vectors $\hat{e}_\alpha$ is $-\frac{1}{3}$, this model is formally equivalent to an AFM Ising model on the trillium lattice. The trillium lattice is a simple cubic structure with a four site basis and a $P2_13$ symmetry featuring a 3D lattice of corner-shared equilateral triangles as pictured in Fig. \ref{figure1}.

 It is interesting to note that none of the spin structures corresponding to the ground state of the AFM Ising model are common to the ground state of the classical AFM Heisenberg model on any corner-shared equilateral triangle lattice in contrast to the case on the pyrochlore lattice. This means that the excellent agreement{\cite{Isakov}} between large-$N$ studies of the AFM Heisenberg model and the local FM Ising model on the pyrochlore lattice is not expected to carry over to studies of corner-shared triangle lattices.

 In contrast to earlier studied spin ice lattices, the AFM Heisenberg model on the trillium lattice is known to order at low temperatures{\cite{Hop,Kee}} despite the geometric frustration inherent to its structure. We will show below that this is not the case for trillium spin ice meaning that the FM local Ising model studied here is the first fully frustrated model to be studied on this structure. The ground state of any corner-shared triangle structure can be simply found by minimizing the Hamiltonian of Eq. {\ref{equation8}} on a triangle and counting the energy per spin. For $J < 0$, this leads to a ground state configuration with two spins out(/in) and one in(/out) with a ground state energy of $\frac{J}{3}$ per spin{\cite{note2}}.

\begin{table}
  \begin{tabular}{ | c | r | l | }
    \hline
    $\alpha$ & $\hat{e}_\alpha$ & Spin Location \\ \hline
    1 & $\frac{1}{\sqrt{3}}(1,1,1)$ & $(u,u,u)$ \\ \hline
    2 & $\frac{1}{\sqrt{3}}(1,-1,-1)$ & $(u+\frac{1}{2},\frac{1}{2}-u,1-u)$ \\ \hline
    3 & $\frac{1}{\sqrt{3}}(-1,1,-1)$ & $(1-u,u+\frac{1}{2},\frac{1}{2}-u)$ \\ \hline
    4 & $ \frac{1}{\sqrt{3}}(-1,-1,1)$ & $(\frac{1}{2}-u,1-u,u+\frac{1}{2})$ \\ \hline
  \end{tabular}
  \caption{Unit vectors $\hat{e}_\alpha$ denoting the local easy axis at each location within a unit cell for four inequivalent sites labelled by $\alpha$. The relative location of each site within the unit cell is given in terms of a crystal parameter $u$.}\label{table1}
\end{table}

\section{Method}

 We have carried out Monte Carlo simulations using the Metropolis algorithm with periodic boundary conditions on cubic lattices of side length $L$ corresponding to $4L^3$ spins. The bulk of our analysis corresponds to a choice of $L = 6$, which contains 864 spin sites, as our results appeared to reach the thermodynamic limit by $L=3$. To ensure that we had reached the thermodynamic limit we considered systems as large as $L = 18$ in our calculation of the residual entropy of this spin system. At each temperature, 10000 Monte Carlo steps were used to equilibrate the system and a further 1000 were used to calculate the averages of physical quantities, where one Monte Carlo step was taken to on average attempt one update per site. Error bars as reported correspond to the standard deviation of our averages over four independent trials. Following the simulated annealing prescription of Kirkpatrick {\it{et al.}}{\cite{kirkpatrick}}, for most of our analysis the temperature of the system was started at $T = 20J$ and reduced by $1$\% at each step. Exact results for the residual entropy were found by counting unique members of the ground state for the tractable periodic lattice sizes $L = 1$ and $L = 2$.

\begin{figure}
\includegraphics[scale=0.25]{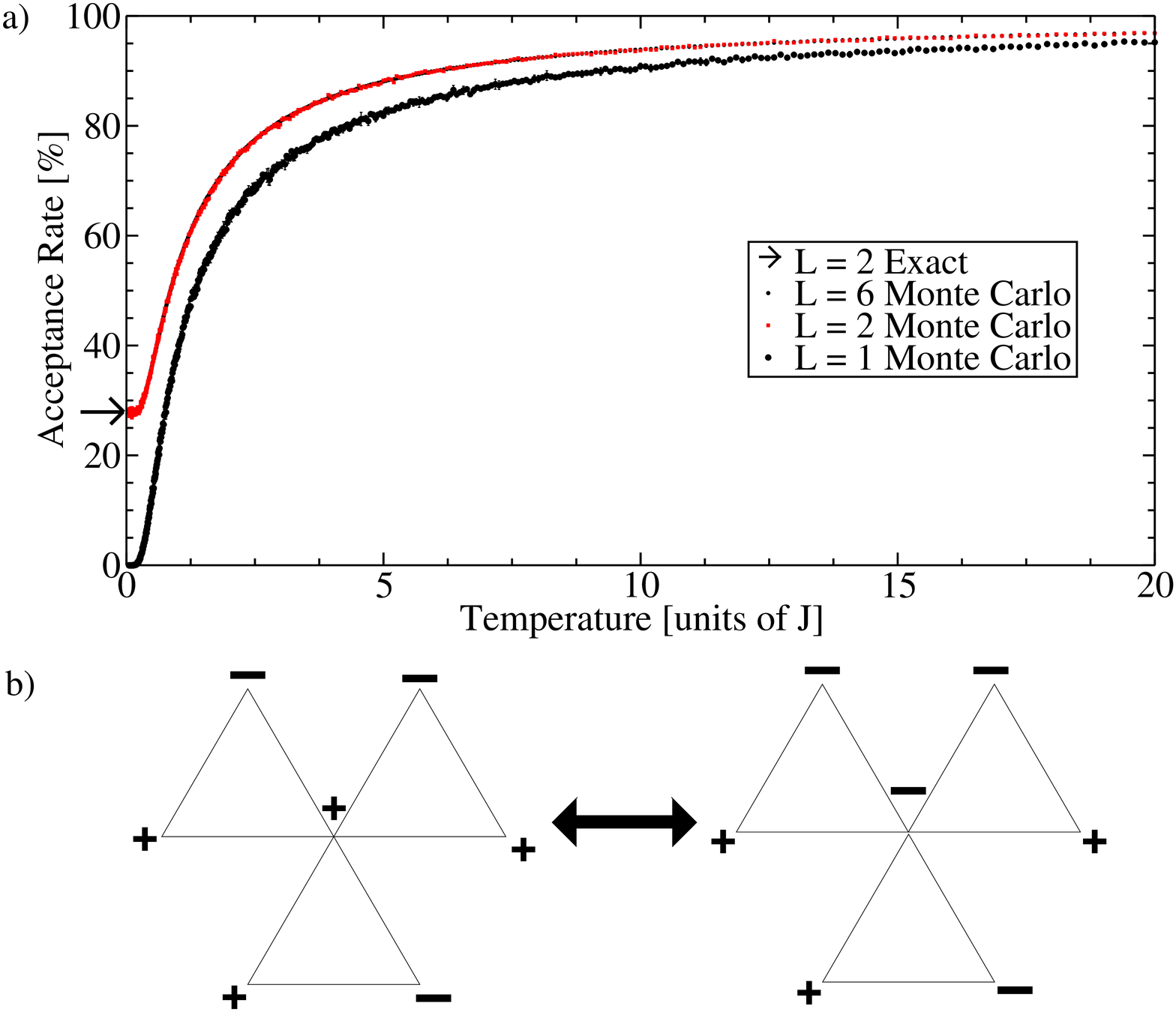}
\caption{(Color online) a) The acceptance rate of the spin flips as a function of the temperature. For $L \geq 2$, below about $T = 0.2|J|$ the acceptance rate asymptotes to the $L = 2$ exact result of 27.8\% as discussed in the text, b) The only zero energy spin fluctuations allowed in the ground state feature spins whose neighbours all pair with opposite signs. Changing the direction of one of the sites changes three triangles from being mostly up to mostly down or vice versa as discussed in Sec. IV. B. {\it{2}}.\label{figure6}}
\end{figure}

\subsection{Acceptance rate}

\begin{figure}
\includegraphics[scale=0.40]{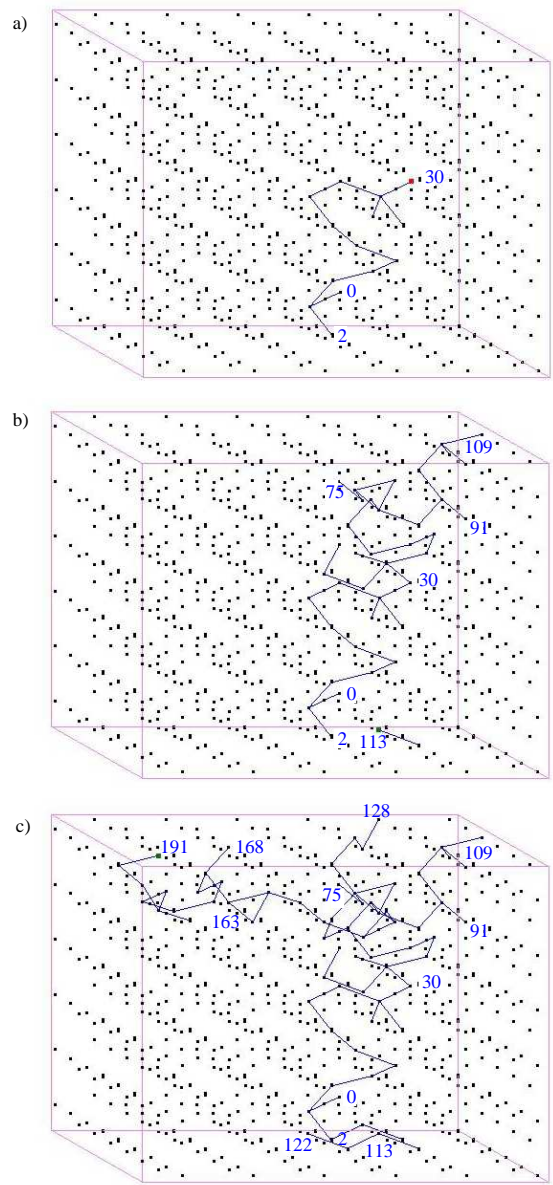}
\caption{(Color online) The evolution of the location of a particular flippable site in time. The path of spin flips which originated at site 0 after a) 30, b) 109, and c) 191 updates. The numbers in the pictures are the number of updates it took for the path to reach that spin site. The number of updates includes updates where the flippable location does not change, instead the same site flips back and forth between its two possible directions.\label{figure7}}
\end{figure}

 As shown in Fig. \ref{figure6}(a) in contrast to studies{\cite{Melko}} of spin ice physics on the tetrahedron-based pyrochlore lattice, at low temperatures on our triangle-based trillium lattice, the acceptance rate of the Monte Carlo simulations does not approach zero except in the case of $L = 1$.  This difference is easily understood in terms of the acceptable ground state spin configurations as shown in Fig. \ref{figure6}(b).  Such flippable sites do not exist on tetrahedron based spin ice lattices.  Counting the total number of ground states for such a 3 triangle configuration which have a flippable spin gives 16 of 54 members of the ground state, slightly above our low temperature asymptote.  Such flippable configurations cannot exist for $L=1$.  A more careful exact treatment for the 32 sites of $L=2$ gives a total of 314874 ground state spin configurations.  Of these 10075968 spin sites, we have numerically found that 2797714 sites are flippable, yielding an exact result of 27.8\% of all sites are flippable, for comparison with the $L=2$ Monte Carlo asymptote of $27.9\%\pm 0.3\%$.  As one can see from Fig. \ref{figure6}(a), there is no discernible difference in the acceptance rate for $L=2$ and $L=6$, indicating that we have quickly reached the thermodynamic limit.

 Lest one be tempted to suggest that for a given simulation one has roughly $73\%$ of the spin sites exhibiting long range magnetic order amidst a background of static flippable spin sites, it is interesting to note that the location of the flippable sites is not fixed. Indeed, even at the lowest temperatures when the system can no longer access spin configurations outside the ground state spin manifold, after waiting a sufficiently long time every site on the lattice will flip. In this sense our model remains disordered (unlike pyrochlore spin ice which becomes stuck in a glassy state\cite{diep2}) and able to equilibrate even to zero temperature. The dynamic process through which flippable sites move around the lattice can be illustrated by following the evolution of the location of one of the flippable sites as shown in Fig. \ref{figure7}. Here we see that it is possible to have neighbouring flippable sites. When a spin flip occurs at site one, there is some probability that a neighbouring spin at site two is now also a flippable site. When this is the case, the next flip could occur either at site one or site two. If the latter, the spin at site 1 may well no longer be flippable, and the location of the flippable site begins to migrate around the lattice. As seen in Fig. \ref{figure7} this migration covers a reasonably large distance in a small number of updates in a manner reminiscent of a random walk or traveling salesman problem.

\section{Results}

\subsection{Heat capacity}

 Within a Monte Carlo simulation one computes the heat capacity as:
\begin{equation}
C_V(T) = \frac{< \! E^2 \! > - < \! E \! >^2}{T^2}\text{,}\label{equation1}
\end{equation}
where $E$ is the energy of a particular spin configuration on the lattice and $T$ is the temperature of the system.

 Magnetically frustrated materials are distinguished by their lack of magnetic order to very low temperature despite the presence of strong spin-spin correlations. In particular, spin ice materials are known to exhibit a soft peak in the heat capacity indicating the onset of the ice rule at low temperatures. At high temperatures there is  little energetic cost for flipping a spin in a disordered system. As more triangles begin to satisfy the ice rule, defects become more isolated forcing the system to pay an energetic cost to flip certain spins. Below the peak essentially all triangles are members of the ground state, with all accepted spin flip processes leaving the system in its ground state, thus paying no energetic cost.

\begin{figure}
\includegraphics[scale=0.28]{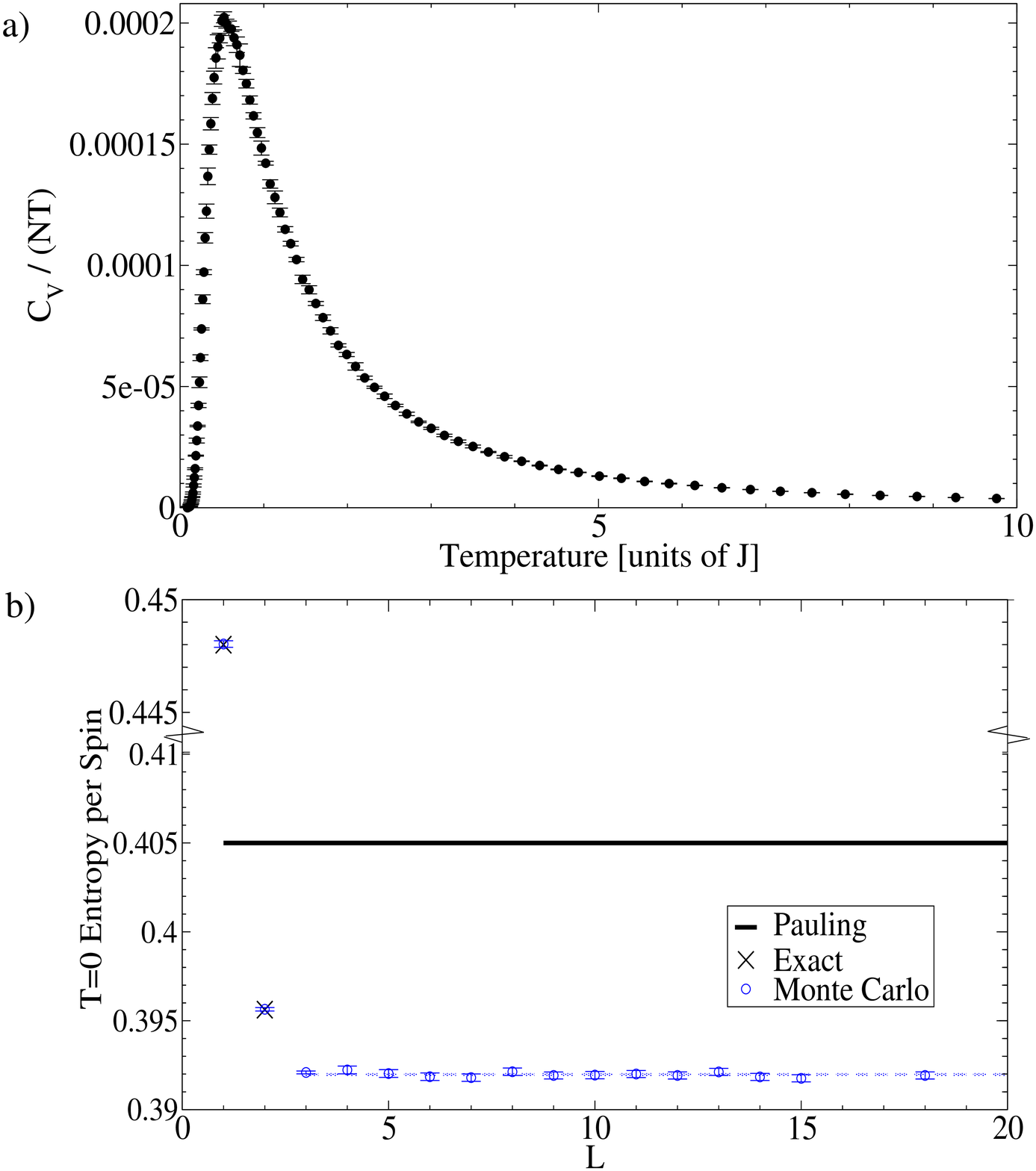}
\caption{(Color online) a) Heat capacity divided by temperature per spin versus temperature in units of the coupling constant $J$ of the system at different temperatures for $L = 6$, b) Residual entropy per spin versus system size, $L$, for a cubic lattice with $4L^3$ sites ($L$ is the number of unit cells in each direction on the lattice). For $L = 1$ and $2$ our Monte Carlo results agree with exact results. For $L \geq 3$ we appear to have reached the thermodynamic limit with an entropy $4\%$ lower than than Pauling's approximation. Note that the region corresponding to the Pauling entropy has been blown up to make our error bars visible, and the $L = 1$ result lies considerably higher.\label{figure3}}
\end{figure}

 In Fig. \ref{figure3}(a) we present the heat capacity per spin divided by temperature as a function of temperature for $L = 6$. We observe a characteristic soft peak in the heat capacity with the ice rule obeyed by the majority of triangles below about $\frac{J}{2}$, and completely obeyed below $T\sim\frac{J}{8}$.


\subsection{Entropy}

 At very high temperature one expects there to be essentially no correlations between the local Ising spins, or $2^N$ equally weighted spin configurations for an entropy per spin of $\ln(2)$. In a majority of materials, as the temperature approaches zero, in accordance with the third law of thermodynamics, the entropy of the system approaches zero. Spin ice materials are a notable exception to this rule, as to very low temperatures they retain a large finite residual entropy. One can quantify the amount of residual entropy at temperature $T$ by subtracting the integrated weight under the $\frac{C_V}{T}$ curve from the high temperature entropy of the system,

\begin{equation}
S(T) = S(T=\infty) - \int_T^\infty \! \frac{C_V(T)}{T} \, dT\text{.}\label{equation2}
\end{equation}

 Here and below we will refer to the entropy per spin for our model. In the $T\rarrow0$ limit, we have a finite residual entropy and correspondingly macroscopic ground state if $\frac{S}{N}(0) = \ln(2) - \int_0^\infty \! \frac{C_V}{T} \, dT > 0$.

\subsubsection{Pauling estimate}

 As noted above, the simplest approximation one can make to the residual entropy of a corner-shared triangle/tetrahedron lattice is to assume, as Pauling did in the context of water ice\cite{Pauling}, that spins beyond the nearest neighboring triangle are uncorrelated. Then each spin site has two degrees of freedom constrained by only a fraction of the possible spin configurations on each triangle/tetrahedron being members of the ground state. For a triangle-based lattice this number is $\frac{6}{8}$. If the triangle assignments are uncorrelated then for each triangle added one adds a factor of $\frac{6}{8}$ to the computation of the total states available to the system. On the hyperkagome and kagome lattices each spin belongs to two triangles, such that there are $N_t=\frac{2N}{3}$ triangles in total. On the trillium lattice each spin belongs to 3 triangles, so there are $N_t=N$ triangles in total. The Pauling degeneracy of the ground state is then $2^N(\frac{6}{8})^{N_t}=2^N(\frac{6}{8})^N=(\frac{3}{2})^N$, hence the Pauling estimate for the entropy per spin is $\frac{S}{N} = \frac{1}{N}\ln(\frac{3}{2})^N=\ln(\frac{3}{2})$.

\subsubsection{Exact}

 While one might expect the Pauling estimate to provide a reasonable approximation to the ground state degeneracy in the thermodynamic limit for uncorrelated triangles, it clearly has some limitations as the number of unit cells becomes small. As an example, if we consider the case $L=1$, the Pauling estimate includes 4 triangles and 4 spins, leading to a non-integral number of distinct ground states, $\frac{81}{6}$. For this case it is not hard to show that all members of the true ground state have $\sigma_{i\alpha}=+1$ for two sites in the unit cell and $\sigma_{i\alpha}=-1$ for the other two. This describes ${{4}\choose{2}}=6$ states, so the exact entropy per spin for the case $L=1$ is $\frac{S}{N}=\frac{1}{4}\ln(6)$, substantially higher than the Pauling estimate.

 The case $L=2$ is still numerically tractable. If we designate the types of triangles available to the ground state by the number of ``positive'' (featuring $\sigma_{i\alpha}=+1$ on two vertices, $\sigma_{i\alpha}=-1$ on the other) and ``negative'' triangles (the converse), we see from Table \ref{table2} that the geometry of the lattice only allows changes in the number of triangles of one type to vary by three triangles if they are to remain in the ground state. Counting all members, we find $314874$ distinct states, much reduced from the Pauling estimate of approximately $431439.88$ states.

\begin{table}
  \begin{tabular}{ | c | c | c | }
    \hline
    Positive Triangles & Negative Triangles & Number \\ \hline
    4 & 28 & 72 \\ \hline
    7 & 25 & 1824 \\ \hline
    10 & 22 & 19680 \\ \hline
    13 & 19 & 76512 \\ \hline
    16 & 16 & 118698 \\ \hline
  \end{tabular}
  \caption{The number of members of the $L=2$ ground state featuring a given number of ``positive'' ($\sigma_{i\alpha}=+1$ on two vertices and $\sigma_{i\alpha}=-1$ on the other) and ``negative'' ($\sigma_{i\alpha}=-1$ on two vertices, $\sigma_{i\alpha}=+1$ on the other) triangles. The table is symmetric in positive $\leftrightarrow$ negative if continued.}\label{table2}
\end{table}

 Although even the case $L=3$ is no longer numerically tractable, the necessity of changing the number of triangles of any type (positive or negative) by three continues from one set of allowed triangle configurations to the next. This may indicate the presence of correlations not present in the Pauling estimate which could account for our results consistently falling somewhat below this approximation. The origin of this change of three triangles appears to be that a ground state configuration, in order to remain in the ground state with a single change, can only change a spin whose neighbors are all paired in opposite signs on their corresponding triangles as illustrated in Fig. {\ref{figure6}}(b).

\subsubsection{Monte Carlo}

 As shown in Fig. \ref{figure3}(b) the entropy calculated by Monte Carlo simulations\cite{note3} agrees well with the exact results available at small $L$, giving a residual entropy per spin at $L=1$ of $0.44803 \pm 0.00015$ (c.f. with $\frac{1}{4}\ln6\simeq0.44794$) and at $L=2$ of $0.39564 \pm 0.00010$ (c.f. with $\frac{1}{32}\ln(314874)\simeq0.395623$). As expected from Pauling's estimate, we find a large finite residual entropy as we approach the thermodynamic limit. For $L=3$ and higher the entropy of trillium spin ice is consistent with $0.3920 \pm 0.0002$, approximately 96\% of the Pauling entropy ($\ln(\frac{3}{2})\approx0.40547$), as one can see from the dashed horizontal line in Fig. \ref{figure3}(b){\cite{noteSim}}.

\subsection{Spin-spin corrlelations}

 In order to calculate the static spin-spin correlations, the static structure factor (${\mathcal{S}}$) of the trillium lattice was found,

\begin{equation}
{\mathcal{S}}(\vec{q}, T) = \frac{1}{N} \displaystyle\sum_{i,j} \! < \! \vec{s_i}(T) \cdot \vec{s_j}(T) \! > e^{i \vec{q} \cdot (\vec{r_i} - \vec{r_j})}\text{.}\label{equation3}
\end{equation}

 The dot product between all pairs of spins on the lattice was averaged in the same way that the heat capacity was, with 10000 Monte Carlo steps used to equilibrate at each temperature, 1000 Monte Carlo steps used to find the average, and a temperature difference between each temperature step of 0.1\%. The dot product between each pair of spins at a specific temperature was stored prior to multiplication by the phase factor $e^{i \vec{q}\cdot(\vec{r_i} - \vec{r_j})}$ where $\vec{r}_i$ denotes the position of the spin within the lattice and $\hbar\vec{q}$ the momentum transfer vector. The static magnetic susceptibility was found as

\begin{equation}
\chi(T) = \frac{{\mathcal{S}}(\vec{q}=\vec{0}, T)}{T} = \frac{1}{N \, T} \displaystyle\sum_{i,j} \! < \! \vec{s_i}(T) \cdot \vec{s_j}(T) \! >\text{.}\label{equation4}
\end{equation}

\subsubsection{Magnetic susceptibility}

\begin{figure}
\includegraphics[scale=0.27]{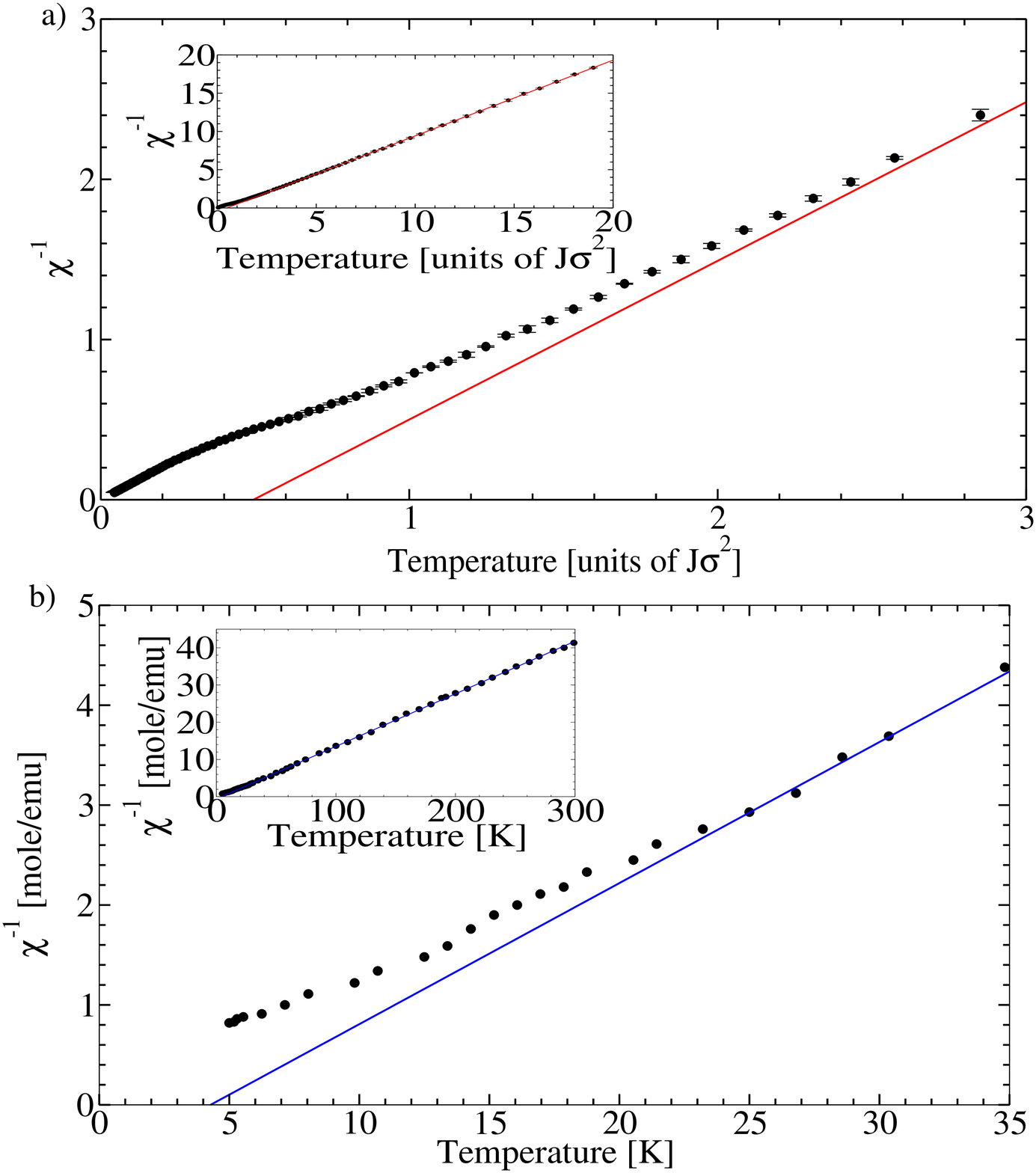}
\caption{(Color online) a) Inverse magnetic susceptibility as a function of temperature in units of the coupling constant J times the square of the spin ($\sigma^2$) from Monte Carlo simulation. b) Inverse molar susceptibility of EuPtSi from D.T. Adroja {\it{et al.}}{\cite{adroja}}. The inset of both graphs shows a linear fit of the high temperature points which was used to define the line that passes through the Curie temperature.\label{figure2}}
\end{figure}

 The temperature dependence of the inverse magnetic susceptibility ($\chi^{-1}$) for $L=6$ is presented in Fig. \ref{figure2}(a). At high temperatures one sees (from the inset) that $\chi^{-1}$ is to a very good approximation\cite{note4} a straight line with a best fit line $\chi^{-1} = (0.991 \pm 0.001)[\frac{1}{J\sigma^2}]T - (0.49 \pm 0.01)$. Below about $T=3|J\sigma^2|$ one sees an upturn of $\chi^{-1}$ relative to this line, with a weak shoulder around $T=\frac{2|J\sigma^2|}{5}$ whereupon $\chi^{-1}$ approaches 0.

\subsubsection{EuPtSi}

 In Fig. \ref{figure2}(b) we show the inverse magnetic susceptibility of EuPtSi as measured by Adroja {\it{et al.}}\cite{adroja}. In this rare earth material, Eu$^{\text{2+}}$ is believed to lie on the trillium lattice (symmetry group of LaIrSi, $P2_13$, and have nearest neighboring Pt and Si atoms lying along the local $<\!111\!>$ axis. This suggests that as in other spin ice materials, single ion anisotropy may act to favour moments along the local $<\!111\!>$ axis. As discussed below, it is not immediately clear why a Eu$^{2+}$ (spin $\frac{7}{2}$, $L=0$) state would experience strong crystal field effects pinning it to a local direction, although some evidence for an analogous favouring of a local transverse magnetization has been presented for equivalent Gd$^{3+}$ ions in Gd$_2$Ti$_2$O$_7$ and Gd$_2$Sn$_2$O$_7$\cite{Glazkov}. Noting the qualitative similarity between Figs. \ref{figure2}(a) and \ref{figure2}(b), it is tempting to speculate that spin ice physics might be responsible for the lack of magnetic order seen to very low temperature in this material. To this end, we have fit the high temperature experimental data with $\chi^{-1}_{cgs} = (0.1412 \pm 0.0005)[\frac{mol}{emu K}]T - (0.61 \pm 0.09)[\frac{mol}{emu}]$\cite{note5}.

\begin{figure*}
\includegraphics[scale=0.24]{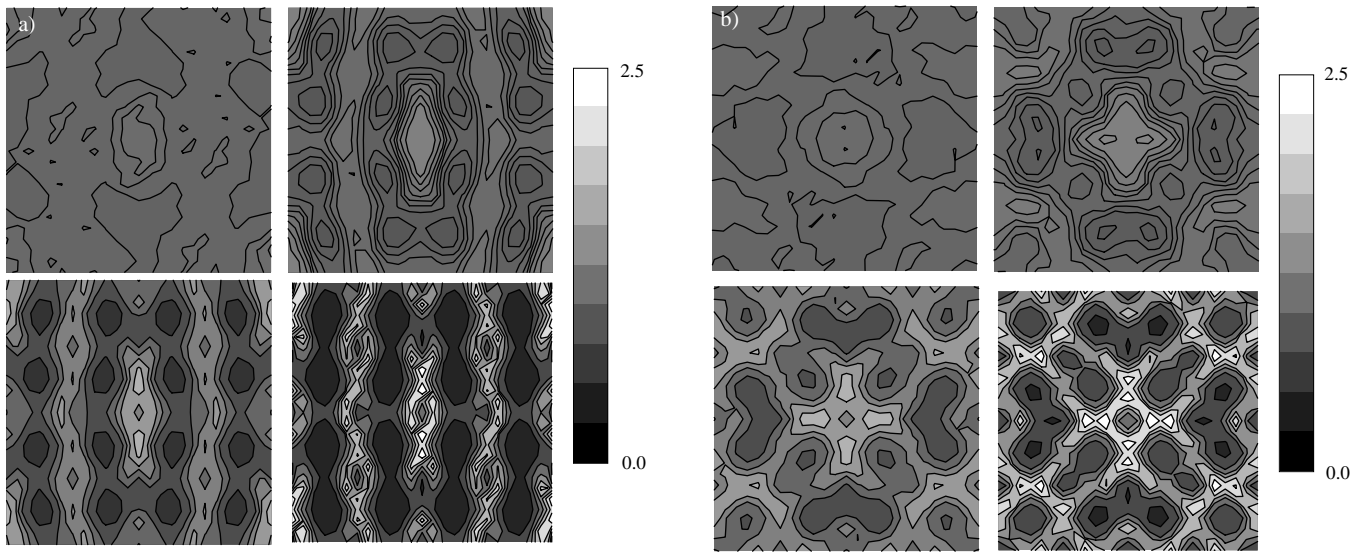}
\caption{Relative intensity of the structure factor in the a) hhk and b) hk0 plane. Temperatures decrease from top left ($10.266J$), to top right ($1.787J$), to bottom left ($0.709J$), to bottom right ($0.0642J$). The axes range from $-\frac{4\pi}{a}$ to $\frac{4\pi}{a}$ where $a$ is the lattice constant with k on the vertical axis and h on the horizontal. The structure factor was calculated using $u=0.138$ as discussed in the text. The relative strength of the correlations is shown on the right of the graphs with white indicating areas of strong correlation and black indicating areas of weak correlation. Although the color scale remains the same for each image, the number of divisions is 7 for the lowest temperature, 10 for $0.709J$, 50 for $1.787J$, and 90 for $10.266J$.\label{figure4}}
\end{figure*}

 To convert this to SI units we need the unit cell length of EuPtSi: $6.436\AA$\cite{adroja}. Noting that $\chi_{SI}=4\pi\chi_{cgs}\Rightarrow\chi^{-1}_{SI}=\frac{1}{4\pi}\chi^{-1}_{cgs}$, and that there are 4 EuPtSi units per unit cell, we need to multiply our expression for $\chi^{-1}$ in $\frac{\text{mol}}{\text{emu}}$ by
$(\frac{1}{4\pi})\frac{VN_A}{4}=\frac{1}{4\pi}\frac{(6.536\times10^{-8}\text{cm})^3(6.022\times10^{23}\text{mol}^{-1})}{4}\simeq3.194\frac{\text{cm}^3}{\text{mol}}$ to arrive at dimensionless $\chi^{-1}$ for comparison with our theory. Here $V$ is the unit cell volume in cm$^{\text{3}}$ and $N_A$ Avogadro's number. This means $\chi^{-1}_{SI,EuPtSi}\simeq(0.451\pm2\times10^{-3})[\frac{1}{K}]T-(2.0\pm0.3)$. From this information, it should be possible to extract an estimate of the coupling constant $J$, given a few assumptions. If Eu$^{\text{2+}}$ takes a high spin configuration, one expects the spin to be $s=\frac{7}{2}$, and orbital angular momentum to be zero, hence one would expect a Lande $g$ factor of $g=2$. If we ignore the discrepancy\cite{notedip} between the Monte Carlo and experimental $\chi^{-1}$ intercepts, we can estimate based on the high temperature slopes of these lines an effective value of $J\sigma^2 \sim -0.455 \pm 0.002 K$, where $\sigma = \frac{7}{2}$.

\subsubsection{Structure factor}

 In Fig. \ref{figure4}(a) and (b) we present the temperature evolution of the static structure factor in the $hhk$ and $hk0$ planes respectively. One sees that correlations gradually increase in strength until entering the spin ice regime. For comparison with previous work we have set the parameter $u=0.138$, the value relevant for the Mn sites of MnSi, as the value of $u$ for EuPtSi is not known. Qualitatively correlations are seen to be quite different in distribution from those seen for the AFM Heisenberg model on the trillium lattice as shown in Fig. 3 of Isakov {\it{et al.}}\cite{Kee}. In particular, one sees a lot of weight inside the box defined by $\{-\frac{\pi}{a},\frac{\pi}{a}\}$ in each coordinate in the spin ice state, while there is very little if any weight for the AFM model. At first the difference between the spin-spin correlations of these models may appear surprising as one is used to the very good agreement of spin correlations seen on the pyrochlore lattice for a local FM Ising model with mean field calculations of an AFM Heisenberg model\cite{Isakov}. However, as explained above this is a simple difference between corner-shared triangle lattices and corner-shared tetrahedral lattices -- that elements of the ground state manifold of an AFM Ising model are not common to the ground state of an AFM Heisenberg model on corner-shared triangles. Put simply, the sum of the spins on a triangle in an AFM Ising model cannot be 0, which is the condition for the ground state of the AFM Heisenberg model on a triangle.



\section{Discussion}

 In light of the recent interest in the creation of deconfined magnetic monopoles on spin ice lattices, it is interesting to consider whether or not deconfined magnetic charges could be present on any of the three corner-shared equilateral triangle spin ice lattices: kagome, hyperkagome, and trillium. Certainly, on each of these lattices, the nature of the ground state dictates that the coarse grained $\vec{\nabla}\cdot\vec{B}=0$ condition responsible for the ``bowties'' seen in polarized neutron scattering experiments{\cite{Fennell}} of Ho$_2$Ti$_2$O$_7$ cannot hold true for all members of a ground state as in its most basic version it relies on the sum of the spins on each triangle/tetrahedron vanishing. While this can in principle occur for AFM Heisenberg spins on a triangle based lattice\cite{hop3}, the ground state of the FM local Ising model cannot satisfy this condition. One might imagine a more coarse-grained sum over several spins could be satisfied for some members of an Ising ground state, with total spin 0, but it seems highly unlikely that such a coarse graining could capture all members. As such, one might not expect to see the "bowtie" structures common to the neutron scattering structure factor of AFM Heisenberg models on the pyrochlore, kagome, and hyperkagome lattices in the spin ice analogs on kagome and hyperkagome lattices. Given this, one might find it surprising that dipolar spin ice apparently does exist on both of these lattices\cite{Chern,Carter}. It is an interesting open question whether dipolar spin ice exists on the trillium lattice, and whether or not the ground state excitations on any of these corner-shared triangle lattices can be thought of in the language of deconfined magnetic charges.

 On the experimental side, it is worth noting that the trillium lattice is a reasonably common magnetic sublattice featuring an itinerant FM (MnSi), a Kondo insulator (FeSi), and several large moment Eu compounds (EuPtSi, EuPdSi, EuPtGe, and EuIrP) not to mention a polar molecule CO which exhibits a residual entropy slightly smaller than $\ln2$ per spin\cite{melhuish}. Of the Eu compounds magnetic order has not been seen to below 4.2K in the first two, to below 2K in the third, with only the fourth seen to ferromagnetically order\cite{pottgen}. To date the first three materials have been classified as paramagnets, but it is our belief the possibility that they are cooperative paramagnets has not been considered. In this context, it may not have been realized that the magnetic sublattice of these materials is actually a frustrating lattice. The presence of a cooperative paramagnetic phase would help explain why each of these materials appears to have a small yet positive Curie temperature.

 While we are struck by the qualitative (and perhaps quantitative) agreement between our model and magnetic susceptibility data of EuPtSi, it is hard to understand why a local FM Ising model would result from a detailed investigation of this material, given that a half-filled spin shell should be spherically symmetric if the high spin state is adopted. We mentioned that the Eu atom lies between Pt and Si atoms along the local $<\!111\!>$ symmetry axis of the crystal, however there are also 2 nearby triangles of Pt and Si atoms centred along this axis which bring the coordination number of the Eu sites up to 20 when the nearest neighbor Eu atoms are considered. Further, the related material EuPtGe is metallic to quite low temperature\cite{Rainer} (the resistivity of EuPtSi has not yet been measured), suggesting that any magnetic correlations that do arise in EuPtSi (if it is metallic) might have a Ruderman-Kittel-Kasuya-Yoshida nature\cite{note6}. Previous work by one of us{\cite{Hop2}} has shown that the extended Heisenberg model with $J_1$ FM and $J_2$, $J_3$ AFM has regions of momentum space which remain nearly degenerate to low temperatures, suggesting that it may be possible for magnetic frustration to play a role in the suppression of the ordering temperature of EuPtSi even if it is not a true realization of the spin-ice physics we have described in this work.

 Lastly, we should mention the possibility that artificial versions of spin ice on the trillium lattice could be created if it turns out that none of the naturally occurring trillium ice candidates provide an immediate physical realization of this model. When kagome spin ice was proposed, Wills {\it{et al.}}\cite{wills} could hardly have predicted that the first realization\cite{notekag} of their model would take the form of lithographically etched mesoscopic islands, whose only magnetic interactions were dipolar in nature. While it is hard to imagine lithographically etching a 3D lattice it is not inconceivable that such technologies could evolve. Likewise, advances in optical lattices have experimentalists claiming\cite{Pertot} that in the near future model spin systems may be able to be built to order, perhaps yielding an alternate path to the realization of the interesting physics of trillium spin ice.

\section{Conclusions}

 In this paper we have predicted the existence of spin ice on the trillium lattice. We have shown that a local FM Ising model on the trillium lattice has a macroscopic residual entropy of $0.3920 \pm 0.0002$ per spin. We have calculated the heat capacity of this model and have seen that it takes the characteristic shape of spin ice on other lattices, with no signs of magnetic order to the lowest temperatures. The magnetic susceptibility at high temperatures appears to follow a Curie Weiss law with a positive Curie temperature before smoothly deviating toward $\chi^{-1}=0$, exhibiting an intriguing weak shoulder at around $\frac{2}{5}|J\sigma^2|$. We have identified a material whose magnetic susceptibility appears to show some of these features. The neutron scattering structure factor of this model has been seen to noticeably differ from the AFM Heisenberg model on this lattice, a feature we expect to be generic of corner-shared triangle lattices. It remains an interesting open problem whether or not the dynamics of spin excitations on this and other triangle-based spin ice materials can be well-described in terms of free magnetic charges, and whether dipolar spin ice exists on the trillium lattice as it does for other spin ice lattices investigated to date.

\vskip1pc
{\it{Acknowledgments:}} It is a pleasure to thank S.V. Isakov for many useful discussions regarding the implementation of our Monte Carlo code. We would like to thank H.-Y Kee for early discussions of this model (Ref. {\onlinecite{note1}}). We also thank Michel Gingras for useful discussions and Matt Enjalran for a careful reading of this manuscript and many helpful suggestions. This work was supported by NSERC(J.H.) and NSERC USRA(T.R.).


\begin{thebibliography}{00}
\bibitem{note7} In water ice, hydrogen bonding between oxygen and hydrogen atoms from neighbouring water molecules is sufficiently strong that each oxygen atom can be thought of as having four hydrogen neighbours. The two hydrogen neighbours forming the water molecule are slightly displaced towards the oxygen atom lying at the centre of a tetrahedron while the other two are slightly displaced towards the neighbouring oxygen atoms.
\bibitem{note2} If one were to study the analogous AFM Ising case ($J > 0$) of such a configuration across a lattice the ground state energy would be $-J$. Note that these energies differ from those of the ground state of the Heisenberg model on the trillium lattice: for $J > 0$ the minimum energy is $-1.5J$ while for $J < 0$ it is $3J$.
\bibitem{note3} For calculations of the entropy, Monte Carlo simulations were run from a starting temperature of $6000|J|$ down to $5.0\times10^{-2}|J|$.
\bibitem{noteSim} Simulations for lattice sizes with $L \leq 10$ were run with a 0.1\% difference between successive temperatures and those with $L > 10$ were run using a 1\% difference between successive temperatures to reduce the computation time for these large lattice sizes. The case with only a single unit cell was additionally run using a 1\% and 0.01\% difference between successive temperatures. Decreasing the change in temperature reduced the size of the error bars in the entropy but the entropies of the three trials were still within error of the exact solution.
\bibitem{notedip} The origin of such a discrepancy could result from slight changes in the intercept and slope that might arise when more realistic interactions are included in our model, such as the dipolar interaction.  Note that the dipolar interaction is crucial to the experimental realization of all known spin ice materials and of approximately this order of magnitude.
\bibitem{note6}  For such large spins it is likely that dipolar interactions play a role. Dzyaloshinskii-Moriya terms may also be relevant on this lattice as it lacks an inversion symmetry.
\end{thebibliography}
\end{document}